\documentclass[preprint,12pt]{elsarticle}
\usepackage{amssymb,amsmath}
\usepackage{hyperref}

\newcommand{\mbf}[1]{\mathbf{#1}}

\begin{document}
\begin{frontmatter}

\title{Phase diagrams of singlet superconducting states with mixed symmetry}
\author{M.~A.~Timirgazin, V.~F.~Gilmutdinov, A.~K.~Arzhnikov}
\address{Udmurt Federal Research Center of Ural Branch of Russian Academy of Sciences, 426000, Russia, Izhevsk, T. Baramzinoy st., 34}

\begin{abstract}
The competition between the singlet superconducting states with $s$- and $d$-wave symmetry of the order parameter is studied within a single-band model with nearest-neighbor attractive interaction. The zero- and finite-temperature ground state phase diagrams are constructed for different ratios between the nearest- and next-nearest-neighbor electron transfer integrals. The mixed $s+id$ pairing state is shown to form in the intermediate region between $s$- and $d$-waves.
The temperature phase transitions between the pure and the mixed pairing state are found. 
\end{abstract}

\end{frontmatter}

\section{Introduction}\label{intro}
High-temperature superconductivity (HTSC) was discovered more than 30 years ago, but many questions concerning the mechanism of this phenomenon still remain open. One of them is the symmetry of the pairing wavefunction. 
It is generally accepted that the cuprate superconductors have the $d_{x^2-y^2}$-wave (hereinafter $d$-wave) pairing symmetry with a superconducting gap $\Delta_\mbf{k}\sim(\cos{k_x}-\cos{k_y})$~\cite{Harlingen95,Tsuei00}, and the iron pnictides have the extended $s$-wave (hereinafter $s$-wave) pairing symmetry with the gap $\Delta_\mbf{k}\sim(\cos{k_x}+\cos{k_y})$~\cite{Stewart11}. 
However, some spectroscopy experiments show that the superconducting gap does not always follow the simplest $d$- or $s$-wave form. For some cuprates the gap may exhibit its maximum value not at the antinodes, as would be the case for the leading harmonic of the $d$-wave form, but at a different location on the Fermi surface (e.\,g.\,in NCCO~\cite{Blumberg02} and LSCO~\cite{Terashima07} compounds).
The results of the ARPES measurements on the A$_x$Fe$_{2-y}$Se$_2$ (A=Ca, K) family~\cite{Zhang11} revealed no hole-pocket at the $\Gamma$ point of the Brillouin zone that argues for the $d$-wave rather than $s$-wave pairing in iron pnictides~\cite{Maier11}. Moreover, the disappearance of electron pockets in the $X$ points of the Brillouin zone in the strong doping regime for K$_x$Ba$_{1-x}$Fe$_2$As$_2$ also points to the $d$-wave form of superconducting gap~\cite{Thomale11}.
Thus, the experimental data often do not give a simple answer to the question of whether $s$- or $d$-wave pairing symmetry is realized, but they are indicative of strong competition and possible mixing between them.

A comprehensive theoretical investigation of competition between the $s$- and $d$-wave pairings was first carried out by Micnas et al.\,in their review \cite{Micnas90}. The stability of superconducting states with different symmetry was studied in the spirit of Bardeen-Cooper-Schrieffer (BCS) theory at certain values of the nearest- and next-nearest transfer electron integrals $t$ and $t'$. The $d$-wave was found to be stable in the region close to the half-filling, while the extended $s$-wave appeared at the band edges. 
The possibility for the gap to have the mixed $s+d$ or $s+id$ symmetry was suggested in Refs.~\cite{Ruckenstein87,Kotliar88}. O'Donovan and Carbotte have considered the general $s+e^{i\Theta}d$ symmetry and found that in the absence of lattice orthorhombic distortions such a state is stable only if $\Theta=\pi/2$ \cite{ODonovan95}. 
The conclusion about the $s+id$ pairing stability, in contrast to the unstable $s+d$ phase, was confirmed later in Ref.~\cite{Musaelian96}.
The temperature phase diagram was constructed for the case of the nearest-neighbor electron hopping in Ref.~\cite{Kuboki01}, and the triplet $p_x+ip_y$ state was found in the intermediate region between the $s$- and $d$-states. 
The authors of Ref.~\cite{Lee09} have considered the pairing symmetry in FeAs-based compounds within the Ginzburg-Landau theory, and have found out that with a decrease in temperature under certain conditions the $s$-wave can go into the $s+id$ state.
Recent investigations carried out in the weak-coupling limit of the Hubbard model have shown that the full phase diagram at $T=0$ includes a rich variety of singlet and triplet superconducting orders, while the $d$-wave remains the ground state in the vicinity of half-filling \cite{Romer15,Kreisel16,Simkovic16}.

Despite a large number of theoretical works devoted to the competition between the $s$- and $d$-wave pairings and their possible mixing, the results are not systematic enough and do not include the phase diagrams for some important cases. We treat the basic mean-field model of superconductivity to construct the temperature phase diagrams for different values of the $t'/t$ ratio. No specific mechanism of superconductivity is considered, but the scenario of antiferromagnetic spin fluctuations is implied, which induces the singlet $s$-, $d$- and $s+id$-wave pairings. 

\section{Model and Methods}\label{model}
We treat the tight-binding model on a square lattice with the Hamiltonian:

\begin{equation}\label{hamj}
{\cal H} = \sum_{j,j',\sigma}t^{}_{j,j'}c^\dag_{j,\sigma}c^{}_{j',\sigma} - \mu\sum_{j,\sigma}c^\dag_{j,\sigma}c^{}_{j,\sigma} - {V_0}\sum_{j,j'} c^\dag_{j,\uparrow}c^\dag_{j',\downarrow}c^{}_{j',\downarrow}c^{}_{j,\uparrow},
\end{equation}
where $t_{j,j'}$ is the transfer integral equal to $-t$ for the nearest neighbors and $t'$ for the next-nearest neighbors, $\mu$ is the chemical potential, $V_0$ is the parameter of the nearest-neighbor attractive interaction ($V_0>0$).
In the mean field approximation, after Fourier transformation and Bogoliubov diagonalization Hamiltonian (\ref{hamj}) takes the form:
\begin{equation}\label{headham}
 {\cal H} = \dfrac{N \Delta^2_0}{V_0} + \sum_\mbf{k}(\xi_\mbf{k}-E_\mbf{k})+\sum_\mbf{k} E_\mbf{k}(\gamma^\dag_{\mbf{k}0}\gamma^{}_{\mbf{k}0}
 +\gamma^\dag_{\mbf{k}1}\gamma^{}_{\mbf{k}1}),
\end{equation}
where $N$ is the number of sites, $\gamma_\mbf{k}$ are new Fermi operators which describe elementary quasi-particle excitations of the superconducting system, $\Delta_0$ is the magnitude of the superconducting order parameter, $\xi_\mbf{k} = \varepsilon_\mbf{k} - \mu$,
$\varepsilon_\mbf{k}=-2t(\cos{k_x}+\cos{k_y})+4t'\cos{k_x}\cos{k_y}$ is the dispersion law for the square lattice, $E_\mbf{k}$ is the quasi-particle excitations spectrum. We define the order parameter in the general $s+id$-wave form:
\begin{equation*}
\Delta_\mbf{k} = \Delta_0[\left(\cos{k_x}+\cos{k_y}\right)\cos{\pi\alpha}+i\left(\cos{k_x}-\cos{k_y}\right)\sin{\pi\alpha}]\equiv
\Delta_0 \eta_{\mbf{k},\alpha},
\end{equation*}
\begin{equation}\label{result}
\Delta_{0} = \dfrac{V_0}{N}\sum_{\mbf{k}} \eta^*_{\mbf{k},\alpha}\langle c_{-\mbf{k},\downarrow} c_{\mbf{k},\uparrow} \rangle.
\end{equation}
In our parametrization the relative magnitudes of the $s$- and $d$-wave gaps can be defined as $\Delta^s_0=\Delta_0\cos{\pi\alpha}$ and $\Delta^d_0=\Delta_0\sin{\pi\alpha}$.
One can see that $\alpha=0$ corresponds to the pure $s$-wave pairing and $\alpha=0.5$ to the pure $d$-wave.
The formation of $s+id$ pairing is usually associated with the time-reversal symmetry breaking (see e.\,g.~\cite{Li91}). In Ref.~\cite{Fogelstrom97} a surface-induced Andreev reflection was suggested to be responsible for violation of the time-reversal symmetry. However we argue that the $s+id$ pairing may not break the time-reversal symmetry, and the Hamiltonian can retain its invariance with respect to the time-reversal operator which in the general case should be determined in a more complicated way than by mere complex conjugation (\ref{appendix}).

For a given $\mu$, $\Delta_0$ can be determined from the self-consistent equation:
\begin{equation}\label{delta_ground} 
\Delta_0=\dfrac{V_0}{N}\sum_{\mbf{k}} |\eta_{\mbf{k},\alpha}|^2 \dfrac{\Delta_0}{2E_\mbf{k}}\tanh\dfrac{E_\mbf{k}}{2T},
\end{equation}

In order to obtain the ground state of the system, the thermodynamic potential $\Omega$ should be minimized with respect to the parameter $\alpha$:
\begin{eqnarray}\label{omega}
 \displaystyle \Omega = \dfrac{N\Delta^2_0}{V_0}+\sum_\mbf{k}\left(\xi_\mbf{k}-E_\mbf{k}\right)-2T\sum_\mbf{k}\ln\left(1+e^{E_\mbf{k}/T}\right).
\end{eqnarray}
Then the electron concentration can be found from the expression:
\begin{equation}\label{n_ground}
n=\dfrac{1}{N}\sum_{\mbf{k},\sigma}\langle c^\dag_{\mbf{k},\sigma} c^{}_{\mbf{k},\sigma}\rangle= 
    \dfrac{1}{N}\sum_\mbf{k}\left[1-\dfrac{\xi_\mbf{k}}{E_\mbf{k}}\tanh\dfrac{E_\mbf{k}}{2T}\right].
\end{equation}
The described procedure is more general in comparison with the traditional method of finding $\Delta^s_0$ and $\Delta^d_0$ from a system of self-consistent equations (see e.\,g.~\cite{ODonovan95}), and for a number of tested points it gives the same results.

\section{Results and Discussion}\label{results}
\begin{figure}[b!]
\center
\includegraphics[width=10cm]{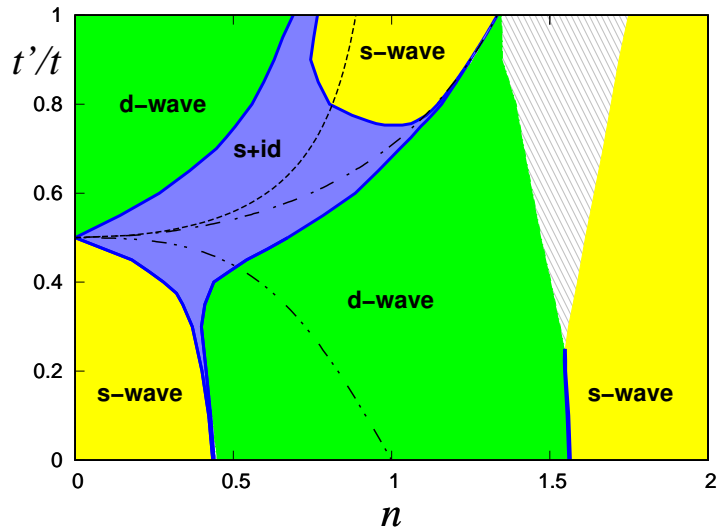}
\caption{The ground-state phase diagram for the square lattice with $T/t=0$ and $V_0/t=1$.
Bold blue lines denote the second-order phase transitions. 
Black dashed lines denote van Hove singularities.
Green color indicates the $d$-wave regions, yellow the $s$-wave, violet the $s+id$.
In the blank shaded area the calculation accuracy does not allow the ground state to be determined, because $\Delta_0$ is too small ($\Delta_0/t<10^{-6}$) .}
\label{groundstate1}
\end{figure}

\begin{figure}[!ht]
\center
\includegraphics[width=10cm]{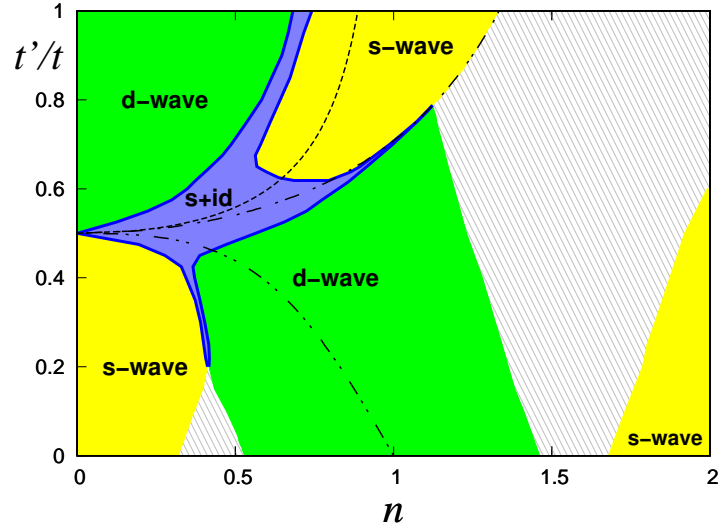}
\caption{The ground-state phase diagram for the square lattice with $T/t=0$ and $V_0/t=0.5$.}
\label{groundstate05}
\end{figure}

\begin{figure}[h!]
\center
\includegraphics[width=10cm]{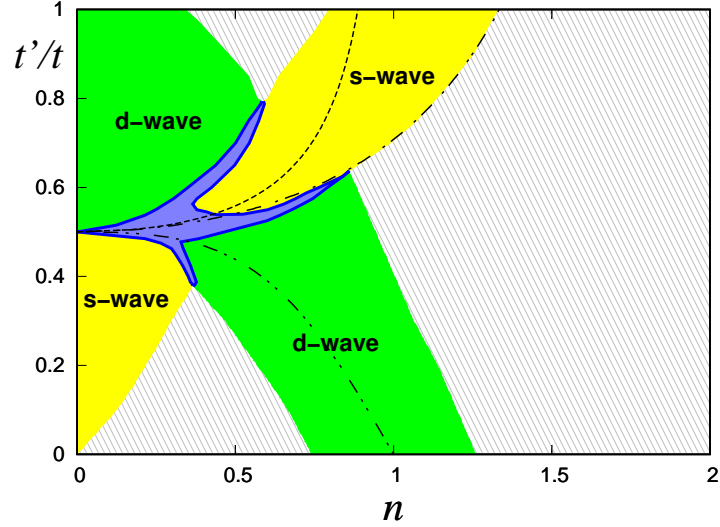}
\caption{The ground-state phase diagram for the square lattice with $T/t=0$ and $V_0/t=0.25$.}
\label{groundstate025}
\end{figure}

The minimization of the Hamiltonian (\ref{omega}) with respect to the mixing parameter $\alpha$ allows one to construct the ground state phase diagrams of the model considered. At the first stage we treat the zero-temperature case. The phase diagrams of the model in terms of $t'/t$ and $n$ are presented in Figs.~\ref{groundstate1}, \ref{groundstate05} and \ref{groundstate025} for $V_0/t=1$, $0.5$ and $0.25$, respectively. 
The phase boundaries between the regions of different wave pairings are determined on a fine grid of parameters $t'/t$ and $\mu$. Integration over the Brillouin zone is performed on an adaptive grid, which allows $\Delta_0/t$ to be calculated with an accuracy of order $10^{-6}$. The regions of $\Delta_0/t<10^{-6}$, indicated by shading on the diagrams, are beyond the precision of our calculations. It should be noted that for arbitrary small $V_0$ the Cooper instability makes the existence of the normal state ($\Delta_0=0$) impossible at $T=0$~\cite{Cooper56}.

For all the considered values of $V_0/t$ the region of mixed $s+id$-wave pairing is presented on the diagram along with the regions of pure $s$- and $d$-wave phases. 
The region occupied by the $s+id$ state decreases with decreasing $V_0/t$. At small values of $t'$ the $s+id$ state region is rather narrow but becomes wider at $t'\gtrsim0.4t$. There are no first order transitions on the diagrams, all the transitions are of the second order with a smooth change of $\Delta_0$ and $\alpha$. At $t'<0.5t$ the pure $d$-wave dominates in the vicinity of half-filling, while the $s$-wave occupies the band edges, as it should be (see e.\,g.~\cite{Micnas90}). At $t'>0.5t$ the picture changes, and the $d$-wave becomes the ground state at the band bottom, while the $s$-wave persists at the band top. At $t'\gtrsim0.8t$ (for $V_0/t=1$) the pure $s$-wave region appears in the vicinity of half-filling. As concerns the pure singlet superconducting states, the diagram (Fig.~\ref{groundstate1}) qualitatively agrees with the diagrams obtained in the small $U$ limit of spin-fluctuation theory~\cite{Romer15,Kreisel16}. This means that the symmetry of the solution appears to be inherent in the model, and is determined by the lattice geometry rather than by the superconductivity mechanism.

\clearpage

\begin{figure}[h!]
\center
\includegraphics[width=10cm]{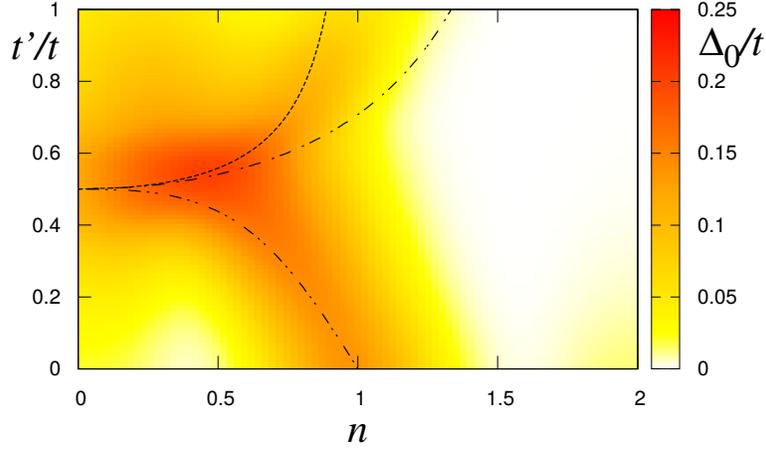}
\caption{Distribution of the superconducting gap magnitude $\Delta_0$. Black lines denote van Hove singularities.}
\label{delta}
\end{figure}

Figure \ref{delta} presents the distribution of $\Delta_0$ over the diagram. 
The maximum of $\Delta_0$ is reached at concentrations which correspond to van Hove singularities of the bare spectrum, which agrees with the van Hove scenario for high-$T_c$ superconductivity~\cite{Markiewicz97}.

\begin{figure}[h!]
\center
\includegraphics[width=10cm]{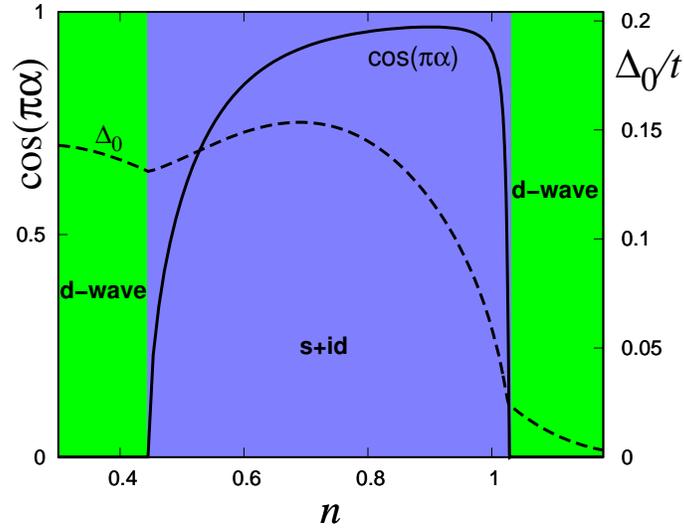}
\caption{Concentration dependence of $\cos{\pi\alpha}$ for $t'=0.7t$ (solid line). Dashed line denotes the $\Delta_0$ magnitude.}
\label{alpha}
\end{figure}

Figure \ref{alpha} shows the concentration dependence of $\alpha$ for $t'=0.7t$. The $s$-wave proportion in the $s+id$-phase increases with concentration, and then decreases rapidly near the boundary with the pure $d$-wave order. At the boundaries between the pure $d$- and mixed $s+id$-phases $\Delta_0$ and $\cos{\pi\alpha}$ change continuously, but their derivatives are discontinuous, which indicates that these transitions are of the second order. This is in agreement with Refs.~\cite{Musaelian96} and~\cite{Liu97}, where second-order phase transitions between the pure and the mixed $s+id$ phase were revealed.

\begin{figure}[b!]
\center
\includegraphics[width=0.49\textwidth]{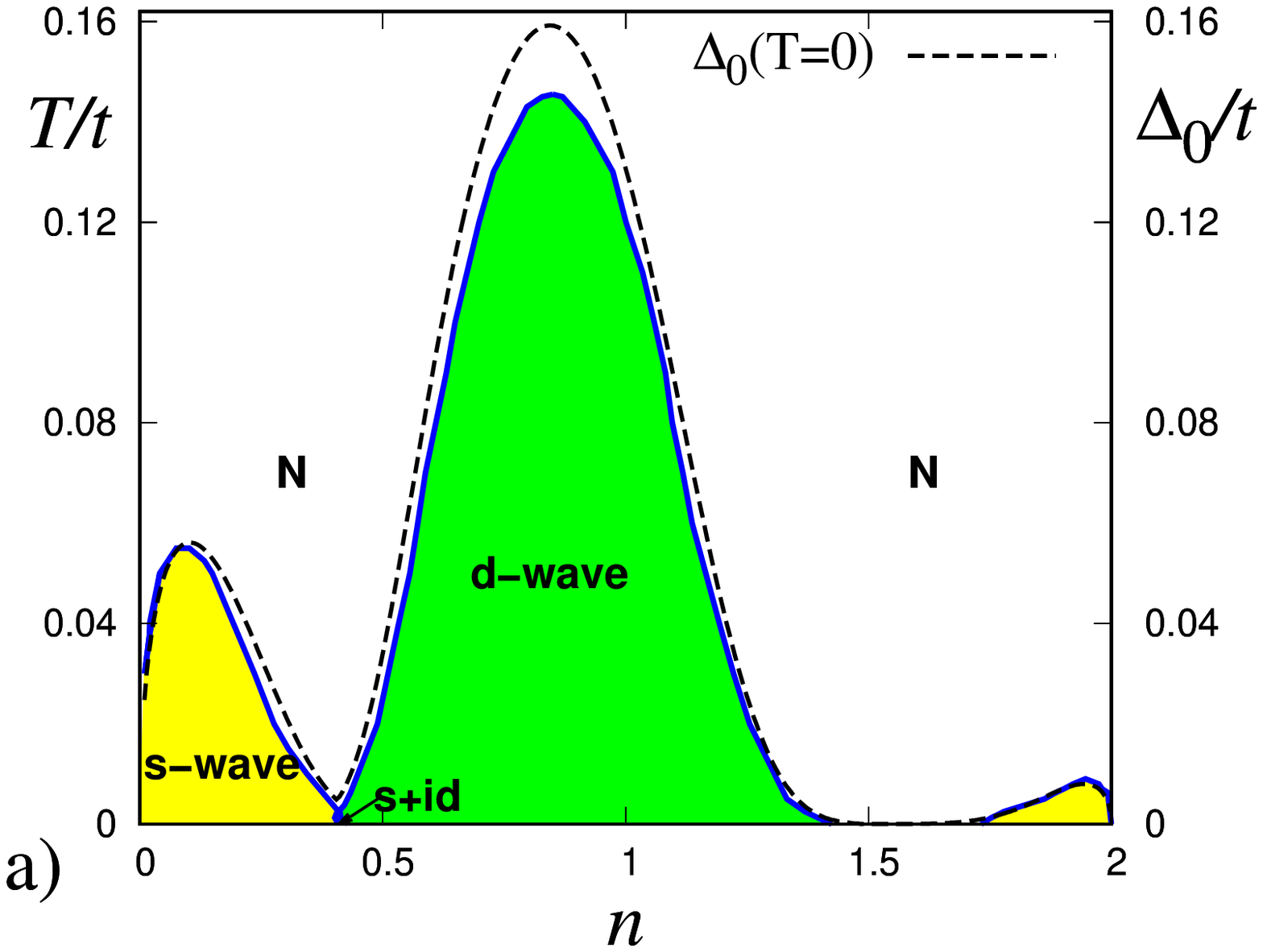}
\includegraphics[width=0.49\textwidth]{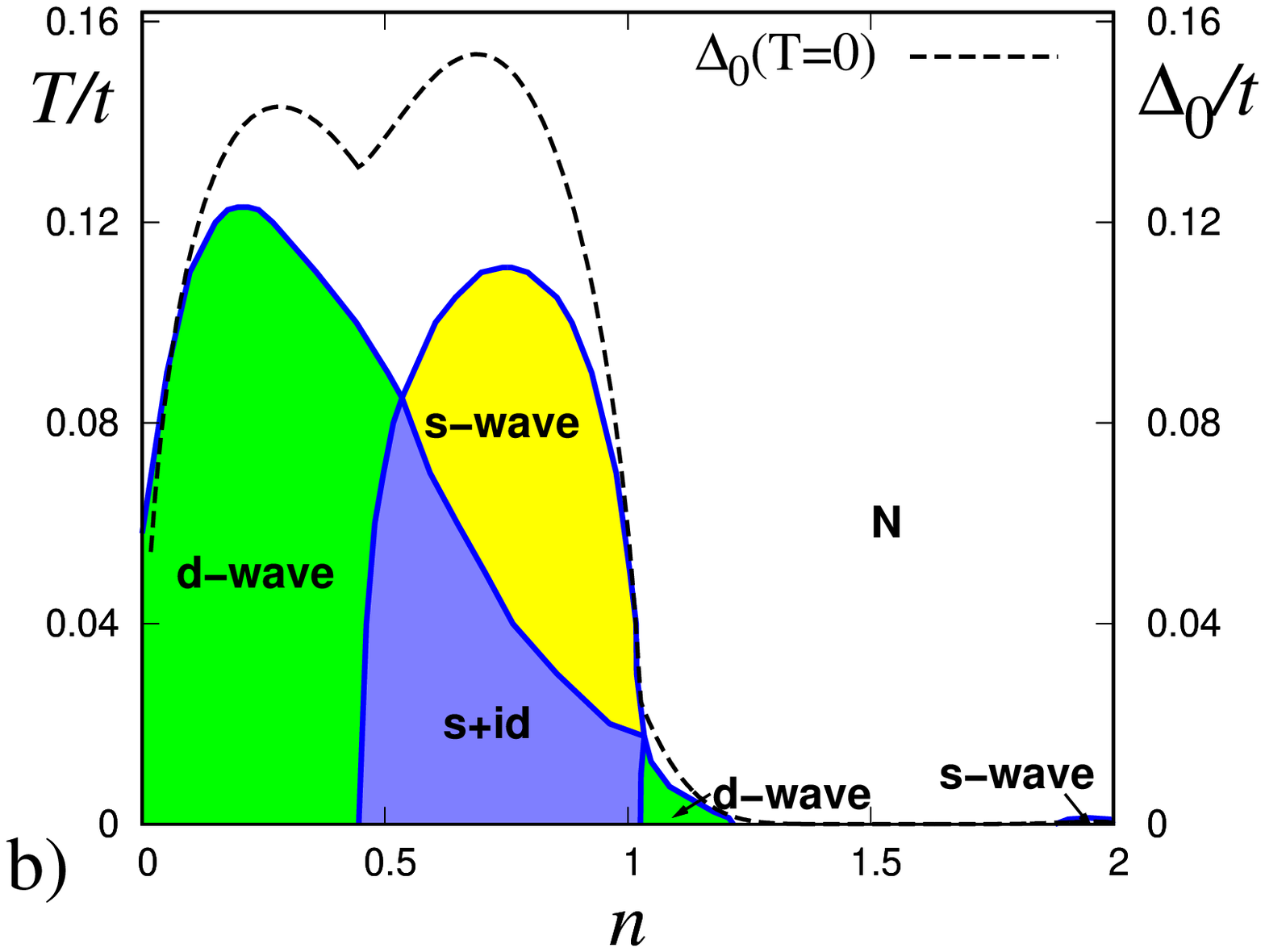}
\caption{Temperature phase diagrams of the model with $V_0/t=1$ for (a) $t'/t=0.2,$, and (b) $t'/t=0.7$. The dashed line shows the value of $\Delta_0$ at $T=0$ for a given $n$.}
\label{temp_1}
\end{figure}

\begin{figure}[h!]
\center
\includegraphics[width=0.49\textwidth]{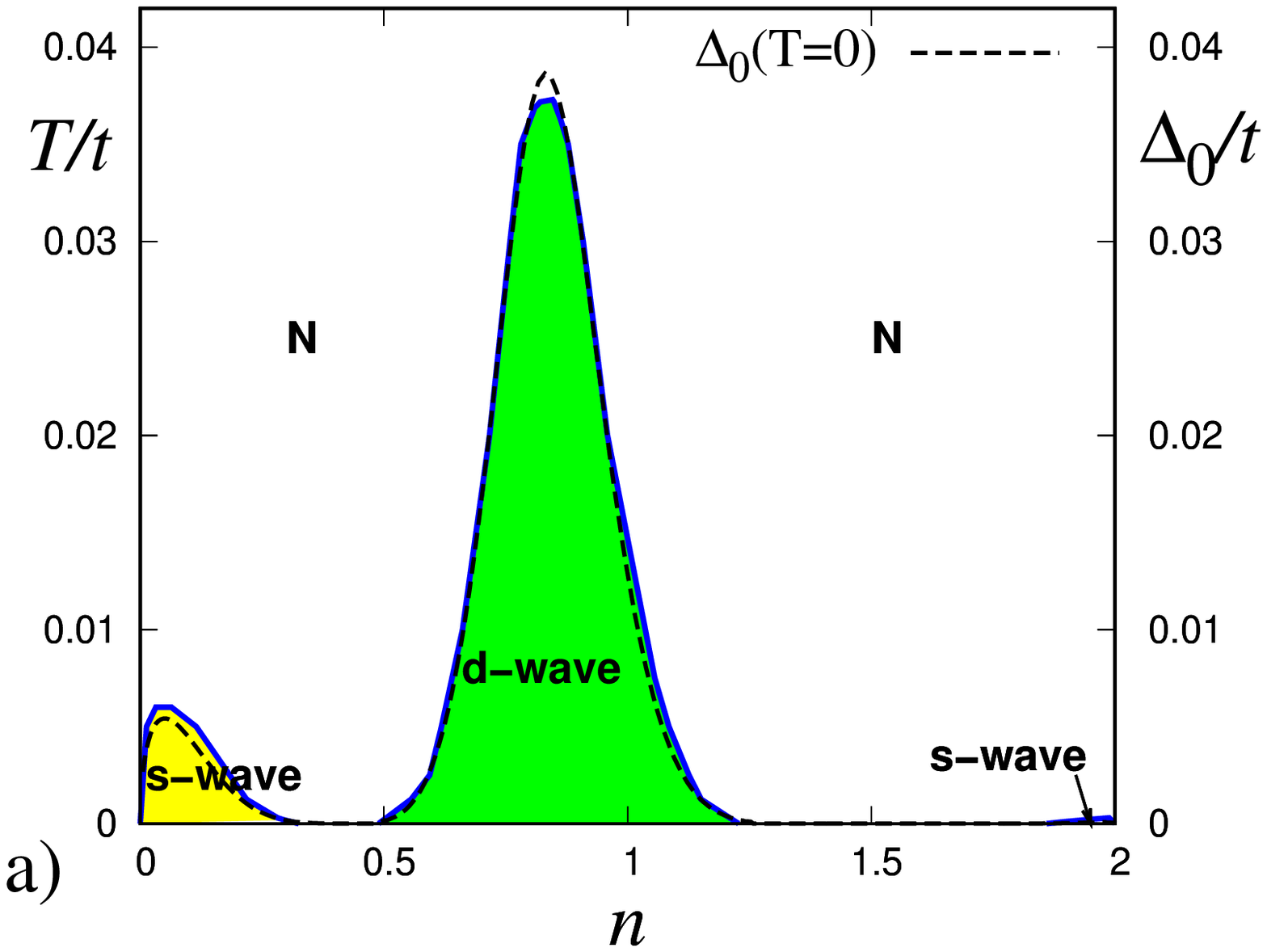}
\includegraphics[width=0.49\textwidth]{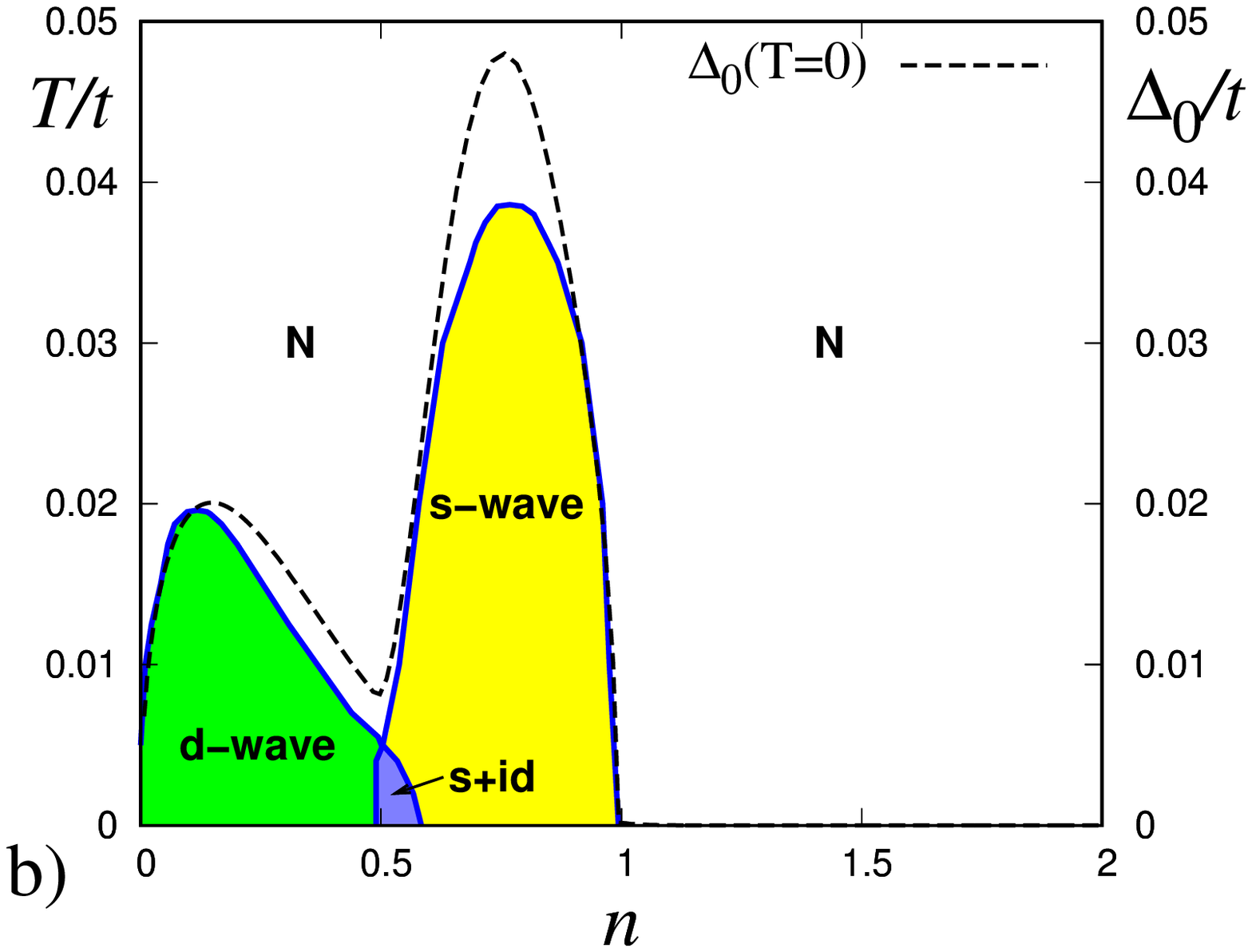}
\caption{Temperature phase diagrams of the model with $V_0/t=0.5$ for (a) $t'/t=0.2$, and (b) $t'/t=0.7$.}
\label{temp_05}
\end{figure}

\begin{figure}[h!]
\center
\includegraphics[width=0.49\textwidth]{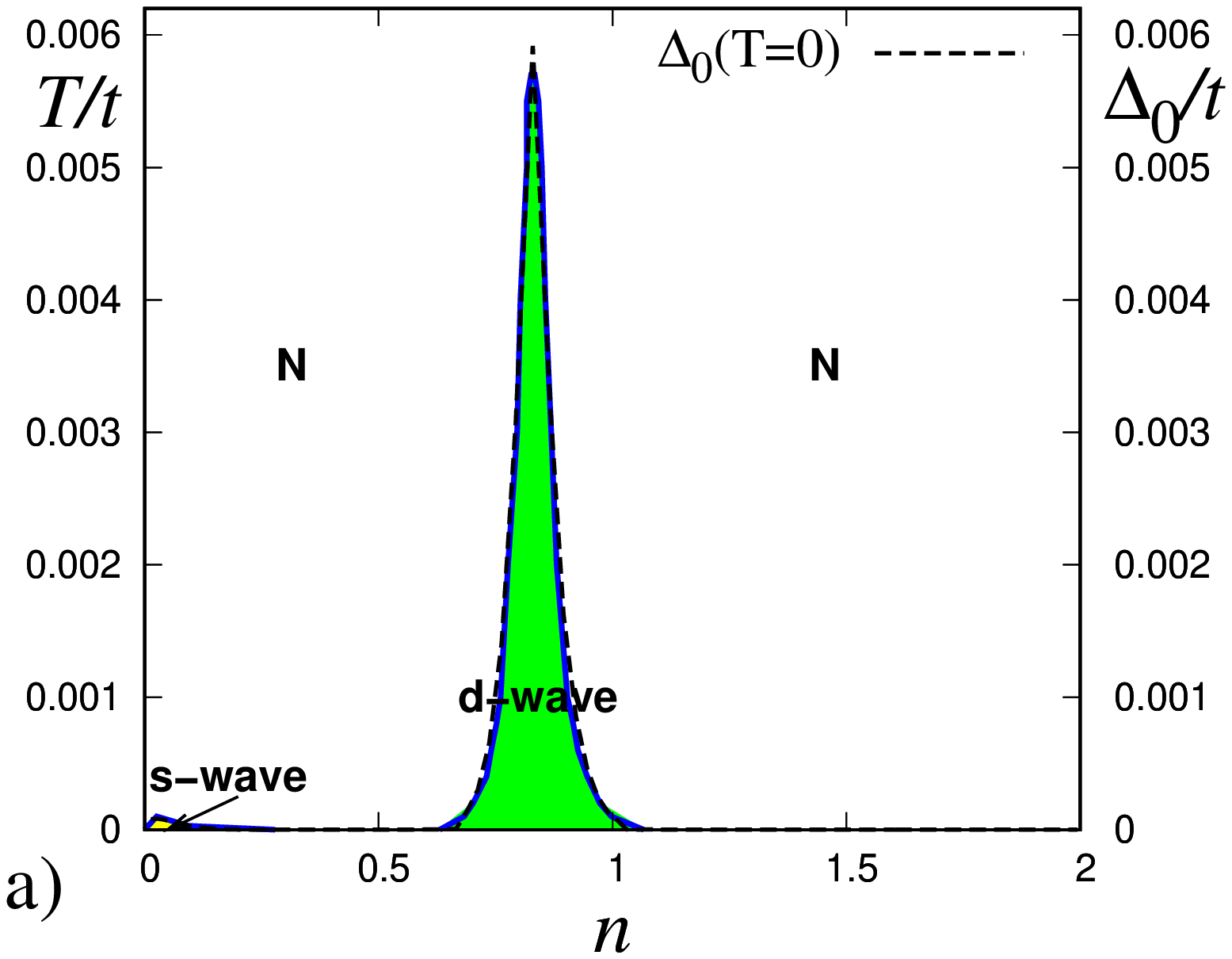}
\includegraphics[width=0.49\textwidth]{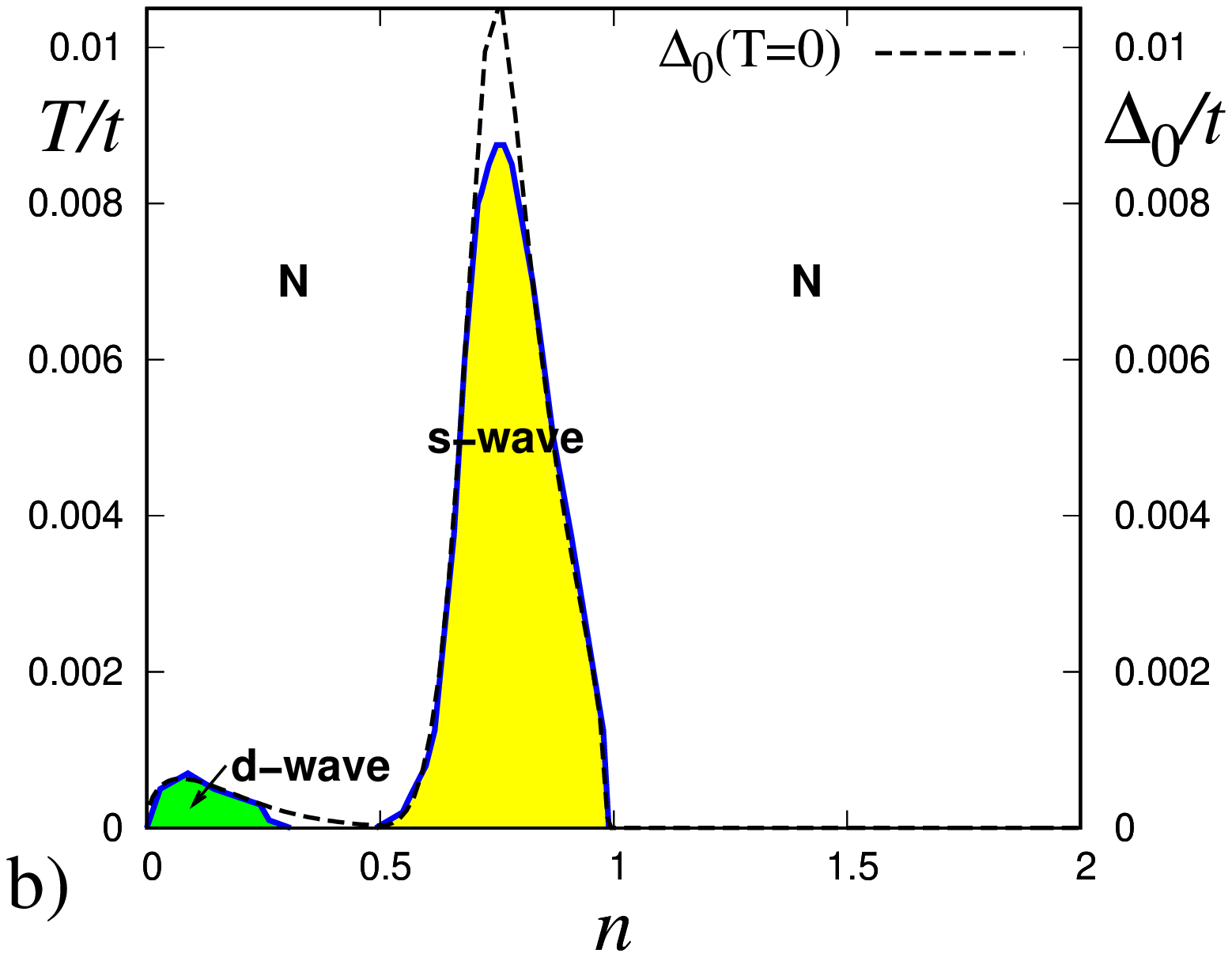}
\caption{Temperature phase diagrams of the model with $V_0/t=0.25$ for (a) $t'/t=0.2$, and (b) $t'/t=0.7$.}
\label{temp_025}
\end{figure}

At the second stage we investigated the finite-temperature case. For all the studied values of $V_0/t$ the temperature diagrams are constructed for the electron hopping parameters $t'/t$ equal to 0.2 and 0.7 : $t'=0.2t$ corresponds to some cuprate superconductors (e.\,g.\,La$_{2-x}$Sr$_x$CuO$_4$~\cite{Hybertsen92}); $t'=0.7t$ corresponds to a wide region of mixed $s+id$ state for $V_0/t=1$ (Fig. \ref{groundstate1}), and to some organic superconductors~\cite{Powell11}. To determine the ground state for the given parameters $T$ and $\mu$ the thermodynamic potential (\ref{omega}) is minimized with respect to $\alpha$. The resulting phase diagrams are presented in Figs. \ref{temp_1}-\ref{temp_025}. In contrast to the diagrams for $T=0$, the normal state appears at high temperature.

The dashed line on the diagrams denotes the dependence of the superconducting gap magnitude $\Delta_0$ at $T=0$ upon the electron concentration. According to the BCS theory this value should be proportional to the critical temperature $T_c$ with the coefficient $\Delta_0/T_c=1.76$. One can see from the figures that the proportional relationship remains valid in our approach, and $1\lesssim\Delta_0/T_c\lesssim1.4$.

The highest $T_c$ value generally corresponds to the phase with the largest $\Delta_0$ at $T=0$, but this is not the case for $V_0=t$, $t'=0.7t$ (Fig. \ref{temp_1}b). The highest $T_c$ is reached for the $d$-wave, while $\Delta_0$ is larger for the $s$-wave. It is interesting that with decreasing $V_0$ this peculiarity vanishes, and $T_c$ for the $d$-wave substantially falls (Figs. \ref{temp_05}b, \ref{temp_025}b). The relative stability of the $s$-wave state for $t'=0.7t$ (and $d$-wave state for $t'=0.2t$) with respect to the interaction decrease can be explained by its vicinity to the van Hove singularity ($n_{vH}\approx0.8$) which favors the superconducting order by the van Hove scenario~\cite{Markiewicz97}.

The diagrams (Figs. \ref{temp_1}b and \ref{temp_05}b) show the possibility of temperature transitions from the pure $s$- or $d$-wave state at high temperatures to the mixed $s+id$ phase at low temperatures. Similar transitions were found in one- and two-band systems and proposed for interpretation of some experimental data for the Fe-based superconductors~\cite{Lee09,Maiti15}. The $s+id$ areas on the diagrams are the largest at $T=0$, and narrow with increasing temperature. No direct transitions from the $s+id$ phase to the normal state are found. 
This result is in accordance with the Landau theory of phase transitions which states that the low-symmetry ordered phase must be transformed by one of the irreducible representations of the point symmetry group of the normal phase. This condition is satisfied for the pure $s$- and $d$-wave phases, but it is not fulfilled for the mixed-symmetry $s+id$ state which is not invariant with respect to complex conjugation.

To summarize, the nearest-neighbor electron attraction can induce superconductivity with $s$-, $d-$ or $s+id$-symmetry. When changing the attraction parameter $V_0$, transfer integral $t'$ and temperature $T$, there can occur different sequences of phase transitions. However, the transition from the $s$- to $d$-wave always goes through the mixed $s+id$ state, the transition from the normal to the superconducting phase is possible only for the pure $s$- and $d$-wave pairing. The $s+id$ state can be realized close to the half-filling, but for reasonable values of the attraction parameter ($V_0<0.25t$) the concentration and temperature ranges of its realization are small and lie far from the half-filling. It should be emphasized that for $t'=0.2t$ which corresponds to HTSC cuprates, the calculated maximum of critical temperature is reached for the hole-doping $\delta\approx0.2$, i.\,e.\,lies in the real optimum doping range. This indicates that the optimum doping is determined by the location of the van Hove singularity rather than by the specific mechanism of the Cooper pairs formation.

\section{Acknowledgements}
This work is supported by Ural Branch of RAS in project 18-2-2-12, and by Russian Foundation
of Basic Research in project of 16-42-180516.

\appendix
\section{}\label{appendix}

The time-reversal operator ${\cal T}$ for the multi-component system is determined by complex conjugation and the action of unitary operator $\cal B$ (see e.\,g.~\cite{Messiah61}):
\begin{equation}\label{time}
 {\cal T}\psi(t)={\cal B}\overline{\psi(-t)},
\end{equation}
where complex conjugation is denoted by a bar.

The commutator with Hamiltonian is
$$
[{\cal T},{\cal H}]\psi(t)={\cal B}\overline{{\cal H}\psi(-t)}-{\cal H B}\overline{\psi(-t)}=({\cal B \overline{H}}-{\cal H B})\overline{\psi(-t)}.
$$

Let ${\cal A}$ be an operator which transforms the Hamiltonian (\ref{hamj}) to the diagonal real operator ${\cal H}'$. Then
$$
{\cal B \overline{H}}-{\cal H B}={\cal B\overline{A}H'}\overline{{\cal A}^{-1}}-{\cal AH'A}^{-1}{\cal B}.
$$
One can see that this expression becomes zero if ${\cal B=A\overline{A}}^{-1}$. Therefore the Hamiltonian is invariant with respect to the time reversal that is determined by the operator (\ref{time}), where ${\cal B}$ for the model considered in the article has the form:
$$
{\cal B}=
\begin{pmatrix}
\dfrac{E_\mathbf{k}(\Delta_\mathbf{k}+\overline{\Delta}_\mathbf{k})-\xi_\mathbf{k}(\Delta_\mathbf{k}-\overline{\Delta}_\mathbf{k})}{2E_\mathbf{k}\overline{\Delta}_\mathbf{k}} & \dfrac{(E_\mathbf{k}-\xi_\mathbf{k})^2}{2E_\mathbf{k}|\Delta_\mathbf{k}|^2}(\Delta_\mathbf{k}-\overline{\Delta}_\mathbf{k}) \\[15 pt]
\dfrac{(E_\mathbf{k}-\xi_\mathbf{k})^2}{2E_\mathbf{k}|\Delta_\mathbf{k}|^2}(\Delta_\mathbf{k}-\overline{\Delta}_\mathbf{k}) & \dfrac{E_\mathbf{k}(\Delta_\mathbf{k}+\overline{\Delta}_\mathbf{k})+\xi_\mathbf{k}(\Delta_\mathbf{k}-\overline{\Delta}_\mathbf{k})}{2E_\mathbf{k}\Delta_\mathbf{k}}
\end{pmatrix}.
$$

\bibliographystyle{elsarticle-num}
\bibliography{mainbbl}

\end{document}